\documentclass[]{mn2e}
\usepackage{graphicx}
\usepackage{epsfig}
\usepackage{amsmath}
\newcommand       \Angstrom     {\,{\rm \AA}}

\newcommand       \cm           {\,{\rm cm}}

\newcommand       \eV           {\,{\rm eV}}
\newcommand       \g            {\,{\rm g}}
\newcommand       \K            {\,{\rm K}}

\newcommand       \s            {\,{\rm s}}

\newcommand       \yr       {\,{\rm yr}}

\newcommand       \Msun     {\,{\rm M_\odot}}
\newcommand       \nH           {n_{\rm H}}
\newcommand     \gtsim  {\lower.5ex\hbox{$\buildrel > \over \sim$}}
\newcommand     \ltsim  {\lower.5ex\hbox{$\buildrel < \over \sim$}}
\newcommand     \simgt  {\lower.5ex\hbox{$\buildrel > \over \sim$}}
\newcommand     \simlt  {\lower.5ex\hbox{$\buildrel < \over \sim$}}

\newcommand       \mum          {\,{\rm \mu m}}
\newcommand       \ppm          {\,{\rm ppm}}

\newcommand       \mH           {m_{\rm H}}
\newcommand       \simali       {\sim\,}

\newcommand	  \ogas        {\left[{\rm O/H}\right]_{\rm gas}}

\newcommand       \nO      {n_{\rm O}}
\newcommand       \vO      {v_{\rm O}}
\newcommand       \mO      {m_{\rm O}}
\newcommand       \MH      {M_{\rm H}}

\newcommand       \msic      {m_{\rm SiC}}
\newcommand       \taucoll   {\tau_{\rm coll}}
\newcommand       \kB   {k_{\rm B}}
\newcommand       \Tgas   {T_{\rm gas}}
\newcommand       \sisic {\left[{\rm Si/H}\right]_{\rm SiC}}
\newcommand       \Nsic {N_{\rm SiC}}
\newcommand       \music {\mu_{\rm SiC}}
\newcommand       \rhosic {\rho_{\rm SiC}}
\newcommand       \dmdtsicpro {\left(dM/dt\right)_{\rm SiC}^{\rm pro}}
\newcommand       \dmdtsicdes {\left(dM/dt\right)_{\rm SiC}^{\rm des}}
\newcommand       \dmdtcpro {\left(dM/dt\right)_{\rm C}^{\rm pro}}

\newcommand{\RNum}[1]{\uppercase\expandafter{\romannumeral #1\relax}}

\title[Where have all the interstellar SiC grains gone?]
        {Where have all the interstellar silicon carbides gone?}
\author[Chen, Xiao, Li, \& Zhou]
{Tao~Chen$^{1,2,5}$\thanks{taochen@kth.se},
C.Y.~Xiao$^{1,5}$\thanks{xiaocunying@bnu.edu.cn},
Aigen~Li$^{3}$\thanks{lia@missouri.edu}, and
C.T.~Zhou$^{4}$\thanks{zcangtao@sztu.edu.cn}\\
        $^1$Department of Astronomy,
                Beijing Normal University,
                Beijing 100875, China\\
        $^2$%School of Engineering Sciences
                Department of Theoretical Chemistry and Biology,
                Royal Institute of Technology,
                10691, Stockholm, Sweden\\
        $^3$Department of Physics and Astronomy,
             University of Missouri,
             Columbia, MO 65211, USA\\
        $^4$College of Engineering Physics,
                  Shenzhen Technology University,
                  Shenzhen 518118, China\\
        $^5$TC and CYX contributed equally.\\          
                  }
\begin{document}
\date{}
\pagerange{\pageref{firstpage}--\pageref{lastpage}} \pubyear{2020}

\maketitle

\label{firstpage}
\begin{abstract}
The detection of the 11.3$\mum$ emission feature
characteristic of the Si--C stretch
in carbon-rich evolved stars
reveals that silicon carbide (SiC) dust grains
are condensed in the outflows of carbon stars.
SiC dust could be a significant constituent of
interstellar dust since it is generally believed that
carbon stars inject a considerable amount of dust
into the interstellar medium (ISM).
The presence of SiC dust in the ISM
is also supported by the identification of presolar
SiC grains of stellar origin in primitive meteorites.
However, the 11.3$\mum$ absorption feature
of SiC has never been seen in the ISM and oxidative
destruction of SiC is often invoked.
In this work we quantitatively explore the destruction
of interstellar SiC dust through oxidation based on
molecular dynamics simulations
and density functional theory calculations.
We find that the reaction of an oxygen
atom with SiC molecules and clusters
is exothermic and could cause CO-loss.
Nevertheless,  even if this is extrapolable to
bulk SiC dust, the destruction rate of SiC dust
through oxidation could still be considerably
smaller than the (currently believed) injection rate
from carbon stars. Therefore, the lack of
the 11.3$\mum$ absorption feature of SiC dust
in the ISM remains a mystery.
A possible solution may lie in
the currently believed stellar injection rate of SiC 
(which may have been overestimated)  
and/or the size of SiC dust
(which may actually be considerably
smaller than submicron in size).

%
%The DFT calculations show that the addition 
%of oxygen atoms are exothermic reactions,
%and the MD simulations demonstrate that
%such reactions weaken or destroy the backbone
%structure of the SiC molecules rapidly. 
%
\end{abstract}

% Accepted 2021 October 28.
% Received 2021 October 13;
%in original form 2021 July 22

\section{Introduction}\label{sec:intro}
Nearly nine decades ago, Wildt (1933)
had already argued that solid silicon carbide
(SiC) grains might form in N-type stars. 
It is now well recognized that SiC solids
are a major dust species, second to amorphous
carbon grains, condensed in the cool atmospheres
of mass-losing, carbon-rich asymptotic giant branch
(AGB) stars (e.g., see Nanni et al.\ 2021).
% ...condensed in stellar outflows...
% ... inject newly-formed dust into the ISM...
This was originally computationally demonstrated
over half a century ago by Friedemann (1969) and Gilman (1969)
based on molecular equilibrium calculations,
and observationally confirmed later
by Treffers \& Cohen (1974) who, by the first time,
detected in two carbon stars, IRC+10216 and IRC+30219,
a broad emission band in between 788 and 973$\cm^{-1}$
attributed to SiC dust. Subsequent spectroscopic observations
have revealed the widespread presence of SiC grains in carbon stars
through the prominent 11.3$\mum$ emission feature
characteristic of the Si--C stretch of SiC solids
(Speck et al.\ 1997, Mutschke et al.\ 1999).\footnote{%
   In principle, essentially all carbon star should make SiC.
   Observationally, SiC dust is harder to detect in redder stars
   as the contrast with amorphous carbon becomes smaller.
   The 11.3$\mum$ SiC feature strength
   with respect to the dust thermal emission continuum
   has been seen to increase with the stellar metallicity
   and show a trend with the stellar mass-loss rate
   (e.g., see Sloan et al.\ 2016, Kramer et al.\ 2019,
   Nanni et al.\ 2021).
   }
In addition, presolar SiC grains of stellar origin
have also been identified in primitive meteorites
based on isotope anomalies
(e.g., see Bernatowicz et al.\ 1987).

% ... the stardust input into the Galaxy ...
% ... estimate the rate of injection of stardust into the ISM 

Although the exact mass fraction of the condensates
in carbon stars which are in the form of SiC
is not precisely known and it depends on stellar mass
and metallicity (Nanni et al.\ 2021),
radiative transfer modeling of the observed
infrared emission of carbon stars has shown
that the mass ratio of SiC to amorphous carbon  
could be as much as $\simali$25\% for the Milky Way
(e.g., see Groenewegen et al.\ 1998)
and $\simali$43\% for the Large Magellanic Cloud
and $\simali$11\% for the Small Magellanic Cloud
(Groenewegen et al.\ 2009, Nanni et al.\ 2019).  
% The SiC feature can also be hidden if you start
% considering coated grains Leisenring et al. (2008).
%
%About 90\% of SiC grains are thought to come from
%low-mass AGB stars of approximately solar metallicity
%(Davis 2011) and SiC accounts for about $\simali$10\%
%of carbon dust of solar and moderately subsolar metallicity
%(Zhukovska \& Henning 2013).
%
Theoretical dust-yield calculations have predicted
that the mass fraction of SiC over the total dust
(i.e., SiC plus amorphous carbon) produced in carbon stars
of an initial mass of 3$\Msun$ around solar metallicity 
is $\simali$25\% (Nanni et al.\ 2013;
also see Zhukovska \& Henning 2013),
although different AGB models produce
different yields at varying metallicity
because of the uncertainties in modelling
the mass-loss and third-dredge up processes
(e.g., see Nanni et al.\ 2013, 2014, 2019;
Ventura et al.\ 2012, 2014, 2018).

Solid grains---SiC and amorphous carbon condensed
in carbon stars as well as silicates and oxides condensed
in oxygen-rich stars will be driven out of the stellar
atmospheres and injected into the interstellar medium (ISM)
by radiation pressure.
If the contribution of carbon stars to the Galactic dust budget
is about comparable to that of oxygen-rich stars as generally
believed (e.g., see Gehrz 1989), then SiC should be a significant
constituent of interstellar dust. However, the nondetection of
the characteristic 11.3$\mum$ feature of SiC in the ISM
puts this at odds (Whittet et al.\ 1990).
%... speculated that interstellar
%SiC may be destroyed by oxidation.
In this work we explore the physical and chemical processes
subjected by SiC in the ISM, with special attention paid to
the destruction of SiC by oxidation. We apply
the Born-Oppenheimer molecular dynamics
(BOMD; Helgaker et al.\ 1990, Uggerud \& Helgaker 1992)
and density functional theory
(DFT; Lee et al.\ 1988, Becke 1992)
to investigate the oxidation reaction pathway
related to SiC grains in the ISM.
The DFT technique is among the most popular and versatile
methods for computational quantum-mechanical modelling
of molecules and clusters. In comparison with other quantum
chemical methods, DFT simplifies the $N$-body problem to
a tractable $3N$ non-interacting system, which makes DFT
more efficient and scalable for large systems.
In BOMD simulations, the energies and forces are
computed from DFT at quantum levels.
Therefore, the BOMD method is more accurate than
classical molecular dynamics in which the energies
and forces are calculated from empirical formulae
or force fields.

% Whittet et al.\ (1990) searched for spectral evidence
%relevant to the possible existence of SiC in the ISM,
%using the line-of-sight to the Galactic center,
%and deduced that the abundance of Si in SiC dust 
%is less than 5\% of that in silicates.
%This is very intriguing, why SiC is not present in the ISM?  
%Where have all the SiC dust grains condensed in C-rich 
%stars and injected into ISM gone? 

%... Thus a grain lifetime and propagation problem is posed.
%... Apparently, it is also necessary to reform and grow grains
%in the ISM, through accretion and coagulation processes,
%in order to explain interstellar dust observations. 

% If carbon stars contribute $\simali$50\% of the total mass
%of dust currently being ejected into the interstellar medium
%by late-type giants, and if much of the available Si in carbon
%star winds is in this form, as has been suggested
%both theoretically and observationally, then SiC should be
%a significant constituent of interstellar dust as well...
%
%While the contribution of AGB stars to the galactic dust budget
%is significant, both in terms of variety and quantity, that due to
%SNe is not yet clear.
%
%In 1969, Friedemann showed that silicon carbide grains
%could condense in the atmospheres of carbon stars and
%then leave the star and become an interstellar dust component,
%although they comprise only a minor fraction of
%the total interstellar dust mass.

\section{Computational Methods}\label{sec:method}
A typical, submicron-sized SiC grain contains hundreds of 
millions of atoms (e.g., there are $\simali$4$\times10^8$
atoms in a spherical SiC grain of radius $a=0.1\mum$).
Even for SiC nanoparticles, it is extremely expensive to study
such systems with {\it ab initio} methods like BOMD and DFT,
requiring tremendous computational power and time.
% which are unavailable even on current supercomputers.
Nevertheless, it is generally believed that SiC grains are
built up via bottom-up processes, starting with small
gas-phase molecules and successive growth to clusters
by molecular additions. Thus, SiC clusters are indispensable
for the formation of SiC grains -- they are the initial states of
SiC grains (e.g., see Gobrecht et al.\ 2017).
Therefore, SiC molecules and clusters could be reasonable
candidates or alternatives of SiC grains for molecular dynamics
modeling. To this end, we consider the most favourable
structures of Si$_{3}$C$_{3}$ and Si$_{12}$C$_{12}$,\footnote{%
    Si$_3$C$_3$ has many isomers and the number of 
    structural isomers for Si$_{12}$C$_{12}$
    gets literally astronomical. Here we are confined
    to the isomers with the lowest energy.
    }
with the former representing a triangle structure
(see Figure~4a of Gobrecht et al.\ 2017,
also see M\"ulh\"auser et al.\ 1993)
with a high surface-to-volume ratio
and the latter exhibiting a spherical structure
(see Figure~16a of Gobrecht et al.\ 2017,
also see Watkins et al.\ 2009) and possessing
a low surface-to-volume ratio.
Si$_{3}$C$_{3}$ is selected because it is
the smallest cluster with a 3D structure
(see Gobrecht et al.\ 2017) and thus allows us
to investigate the reactions for the low-end cluster.
Si$_{12}$C$_{12}$ is selected because of its bucky-like
symmetrical structure which minimizes the considerations
of the incident directions of oxygen atoms.
Si$_{16}$C$_{16}$ is also symmetrical in structure
but it is computationally more expensive
(see Gobrecht et al.\ 2017).

% Nevertheless, SiC grains form through molecular growth
% of SiC molecules, viz., SiC molecules may mark a transition
% from a quantum-confined molecule regime to a crystalline,
% solid bulk-material \citep{gobrecht2017nucleation}.

To evaluate the stability or reactivity of
the studied molecule or cluster,
we calculate the binding and dissociation energies
and transition barriers for each target system,
using DFT as implemented in the Gaussian16 package
(Frisch et al.\ 2016). All structures are optimized
to their ground states using the 6-311++G(2d,p)
basis set. Here the Slater-type atomic orbitals
(AOs) are described by ``basis functions''
and each basis function is described by
a sum of several Gaussian functions.
%
%More specifically, in the 6-311++G(2d,p) basis set,
%  ``6'' indicates that every core AO is described by one basis
%     function which is expressed as a sum of six ``primitive''
%     Gaussian functions; ``-311'' refers to a ``triply-split valence''
%     basis set (three basis functions for every valence AO)
%     and it is primarily this feature that allows
%     the electron density to adopt the best radial
%     distribution for any given bonding situation;    
%     the augmentation of a basis set with diffuse functions
%     is indicated by ``+'' signs, where the first ``+'' refers
%     to heavy atoms and the second ``+'' refers to lighter atoms;
%     and (2d,p) represents a 2d-type Cartesian-Gaussian polarization
%     for heavy atoms and one p-type Cartesian-Gaussian polarisation
%     for light atoms.
% 
To take the intermolecular forces into account,
the D3 version of Grimme's dispersion
with Becke-Johnson damping (Grimme et al.\ 2011)
is included in the calculations.
Here, ``D3'' denotes a function of interatomic
distances, which contains adjustable parameters
that are fitted to the conformational and interaction
energies computed using high-level methods.
  % such as CCSD(T) or CBS. The fitting is done for
  % a given functional. D and D2 energy corrections
  % consider all pairs of atoms while D3 also considers
  % triplets of atoms to account for three-body effects.

The vibrational frequencies are calculated
under the harmonic oscillator approximation
for the optimized geometries.
The transition states are estimated based
on our molecular dynamics simulations
followed by relaxed scanning
(with geometry optimization at each point)
on the potential energy surface (PES).
The transition barrier is calculated
as the energy difference between
the ground states of the reactant
and the transition state.
For reactions without transition state,
the dissociation energy is computed
as the energy difference between
the ground states of the reactant and the product.
The intermolecular forces are taken into
account in all the calculations.
The dynamical processes are simulated
using BOMD (Helgaker et al.\ 1990,
Uggerud \& Helgaker 1992) as implemented
in the Gaussian16 package (Frisch et al.\ 2016).
The B3LYP hybrid functional in combination
with the 6-31G(d,p) basis set is utilized
for the BOMD simulations.
For all these simulations, again,
the D3 version of Grimme's dispersion with
Becke-Johnson damping (Grimme et al.\ 2011)
is also included.
%

%{\bf We note that the vibrational frequencies are calculated
%  based on the harmonic oscillator approximations,
%  i.e., no anharmonic effects are considered.
%  In addition, the empirical scaling factor to compensate
%  for the vibrational anharmonicity and incomplete descriptions
%  of electron correlation is set to 1.0 instead of 0.9877
%  as previously been used. We find that such a modification
%  does not alter the dissociation energies or transition state
%energies significantly ($<$\,1\%). 

%%% Figure 1 %%%
%\begin{figure*}
%\begin{center}
%\includegraphics[width=0.5\textwidth]{fig1.eps}
%\caption{\footnotesize
%               \label{fig:targets}
%               Small silicon carbide molecules
%               studied in this work.
%               Gray balls represent carbon atoms
%               and yellow balls correspond to
%               silicon atoms. Si$_{3}$C$_{3}$ exhibits
%               a triangle structure with a high
%               surface-to-volume ratio,
%               while Si$_{12}$C$_{12}$ shows
%               a highly symmetric tetrahedral
%               structure (symmetry group $T_h$)
%               with a lower surface-to-volume ratio.
%               }
%\end{center}
%\end{figure*}
%%% Figure 1 %%%

\section{Results}\label{sec:results}
Over 30 years ago, Whittet et al.\ (1990) searched for
the 11.3$\mum$ absorption band of SiC in the diffuse ISM
along the line of sight toward the Galactic center.
The lack of any spectral evidence for solid SiC led
Whittet et al.\ (1990) to place an upper limit on
the abundance of silicon relative to hydrogen (i.e., Si/H)
in SiC dust to be at most $\simali$5\% of that in silicate dust.
They speculated that SiC grains could be destroyed in the ISM
by oxidation. Motivated by this, we first model the interaction
between a Si$_{3}$C$_{3}$ molecule and an incident oxygen atom.
This applies to the diffuse ISM where atomic oxygen
is the dominant form of gaseous oxygen.
The oxidation of SiC solids by O$_2$ has been extensively
studied experimentally (e.g., see Ervin 1958) and is not
relevant since there is little O$_2$ in the ISM
(e.g., see Larsson et al.\ 2007).
In dense molecular clouds where oxygen is mostly locked
up in water ice, SiC dust, if present, is expected to be coated
by a layer of water ice and oxidation is unlikely
since the reaction of SiC with water to produce SiO$_2$ and
methane only takes place at a temperature of
above 1300$^{\rm o}$C (e.g., see Park et al.\ 2014).
%

% The reaction system can be described by an incident atom
% (O or H with a temperature of 100 K) colliding with a cold
% SiC grain ($\simali$20 K). The cross section of the incident
% atom (O or H) is insignificant in comparison to the SiC grain
% (containing hundreds of thousands atoms) which only influence
% a few atoms in the SiC grain, i.e., the reactions only take place
%in the area near the incident atom.

%%% Figure 1 %%%
\begin{figure*}
\begin{center}
\includegraphics[width=1.0\textwidth]{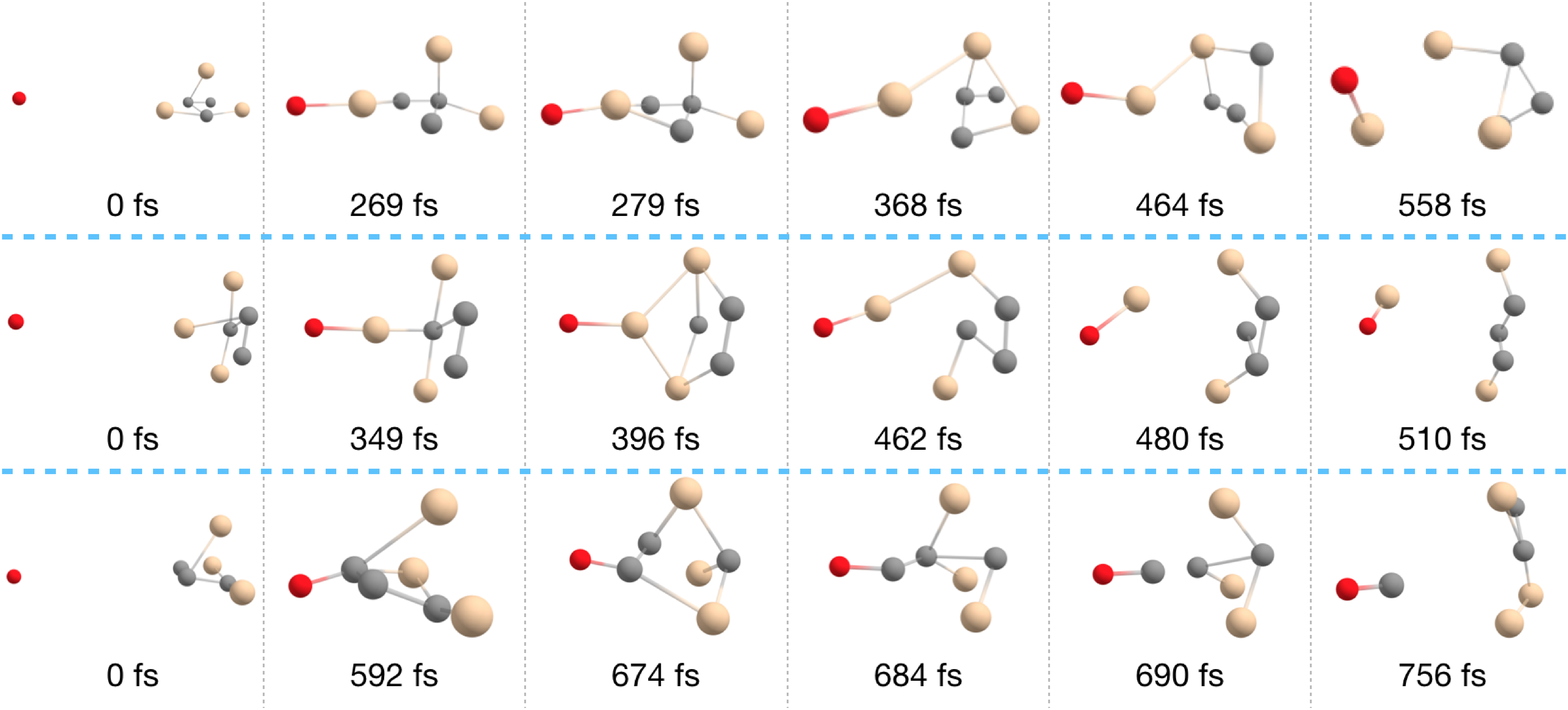}
\caption{%\footnotesize
            \label{fig:snapshot1}
            Snapshots from molecular dynamics simulations
            for collisions between an oxygen atom (red ball)
            and a Si$_{3}$C$_{3}$ molecule
            (in which gray balls represent carbon atoms
             and yellow balls correspond to silicon atoms),
           with the oxygen atom arriving from
           three different directions.
            The actual time in the simulation is shown
            beneath each molecular structure.
            SiO-loss (upper and middle panels) and
            CO-loss (bottom panel) can be seen
            on the rightmost panels.                
            }
\end{center}
\end{figure*}
%%% Figure 1 %%%

As the oxygen atom is much smaller than
a SiC cluster or grain, the incident oxygen atom
would only influence a few atoms
in the target SiC cluster or grain,
i.e., the reaction would only take place
in an area near the incident atom and
depends on the incident direction. 
Figure~\ref{fig:snapshot1} shows the BOMD simulations 
for collisions between a Si$_{3}$C$_{3}$ molecule and
an incident oxygen atom arriving from three different
directions. The simulations demonstrate that, due to
the impact of an oxygen atom, CO and SiO molecules
can be released rapidly from the molecule within 1\,ps.
The fragmentation products are highly dependent on
the incoming direction of the oxygen atom: 
if the oxygen atom first collides with a carbon atom,
then a CO molecule is formed and ejected, 
while a SiO-loss would be triggered
if the incident oxygen atom first hits a silicon atom.

\begin{figure*}
\begin{center}
\includegraphics[width=1.0\textwidth]{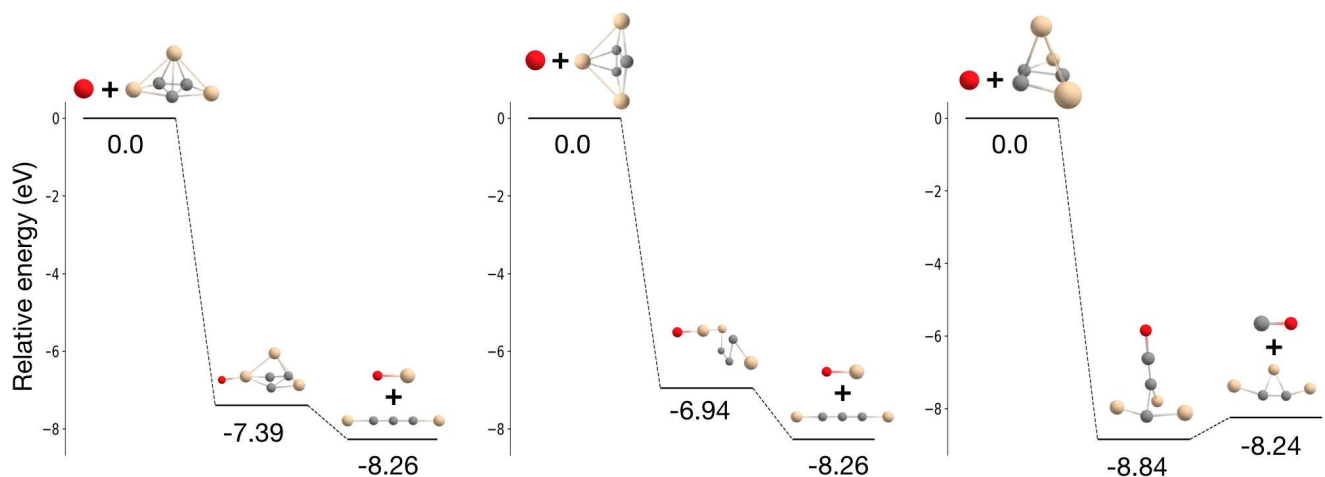}
\caption{%\footnotesize
          \label{fig:Ediagram1}
          Calculated binding and dissociation energies for reactions 
          between a Si$_{3}$C$_{3}$ molecule and an incident oxygen
          atom arriving at three different positions.                
          }
\end{center}
\end{figure*}

To better understand the reaction mechanism,
DFT calculations for the three reaction scenarios
as illustrated in Figure~\ref{fig:snapshot1} are performed
and shown in Figure~\ref{fig:Ediagram1}.
The corresponding binding energies for
O\,+\,Si$_{3}$C$_{3}$ and dissociation energies
for Si$_{3}$C$_{3}$O--CO or Si$_{3}$C$_{3}$O--SiO 
are illustrated in Figure~\ref{fig:Ediagram1} as well.
Most prominently, the reactions for absorbing an oxygen
atom to a  Si$_{3}$C$_{3}$ molecule are {\it exothermic},
which release $\simali$7--9$\eV$ energy,
depending on the incoming direction of
the incident oxygen atom.
This is not surprising since association reactions
     of the form A\,+\,B\,$\rightarrow$\,AB are usually
     exothermic, as the number of bonds is increased
     in the product.
The energy generated from these reactions
will traverse through all of the vibrational modes
of the molecule, and the chemical bonds
with lower binding energies will be dissociated.
Moreover, for the Si$_{3}$C$_{3}$O system
SiO-loss is also an exothermic reaction,
which releases $\simali$1$\eV$ energy.
In contrast, CO has a binding energy of $\simali$0.6$\eV$
and therefore it requires $\simali$0.6$\eV$ to eject CO
from the Si$_{3}$C$_{3}$O system.
However, with the much larger amount of energy
($\simali$7--9$\eV$) gained from the first step,
such a low binding energy can be easily crossed,
leading to a rapid CO-loss.

%%% Figure 2 %%%
%%% Figure 2 %%%

\begin{figure*}
\begin{center}
\includegraphics[width=1.0\textwidth]{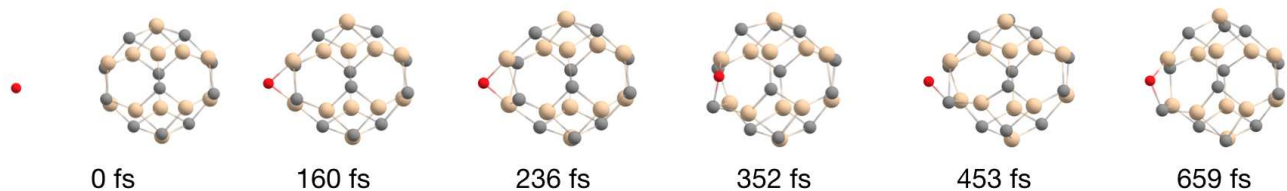}
\caption{%\footnotesize
            \label{fig:snapshot2}
            Snapshots from molecular dynamics simulations
            for collisions between an oxygen atom (red ball)
            and a Si$_{12}$C$_{12}$ molecule.
            Due to the high symmetry of Si$_{12}$C$_{12}$,
            only one incident direction is simulated for
            the oxygen atom.
            }
\end{center}
\end{figure*}

\begin{figure}
\begin{center}
  \includegraphics[width=1.0\columnwidth]{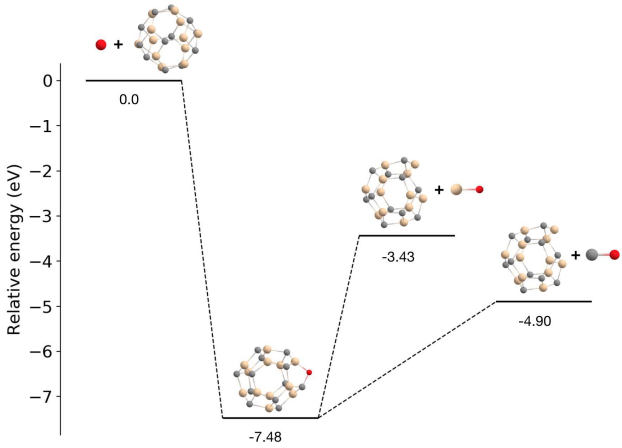}
\caption{%\footnotesize
  \label{fig:Ediagram2}
          Calculated binding and dissociation energies for reactions 
          between a Si$_{12}$C$_{12}$ molecule and an incident oxygen
          atom.
          }
\label{fig:fig4}
\end{center}
\end{figure}

The collisions between an oxygen atom
with larger molecules, e.g., Si$_{12}$C$_{12}$, 
demonstrate a rather different scenario.
Figure~\ref{fig:snapshot2} shows the BOMD
simulations for collisions between an oxygen
atom and a Si$_{12}$C$_{12}$ molecule.
No fragmentation is found, instead, the oxygen atom
is captured by the Si$_{12}$C$_{12}$ molecule.
The symmetry of the molecule is broken
due to the incoming oxygen atom, which 
forms two new covalent bonds, one connected
with a carbon atom and the other one connected
with a silicon atom. The DFT calculations show that,
as illustrated in Figure~\ref{fig:Ediagram2}, 
the absorption of an oxygen atom is still
an exothermic reaction,
which releases $\simali$7.5$\eV$ energy.
However, the dissociation of CO and SiO requires
$\simali$2.6$\eV$ and $\simali$4.0$\eV$, respectively. 
Due to its larger size, Si$_{12}$C$_{12}$ could redistribute
the excess energy across its degrees of freedom
to avoid fragmentation. 

\begin{figure*}
\begin{center}
\includegraphics[width=1.0\textwidth]{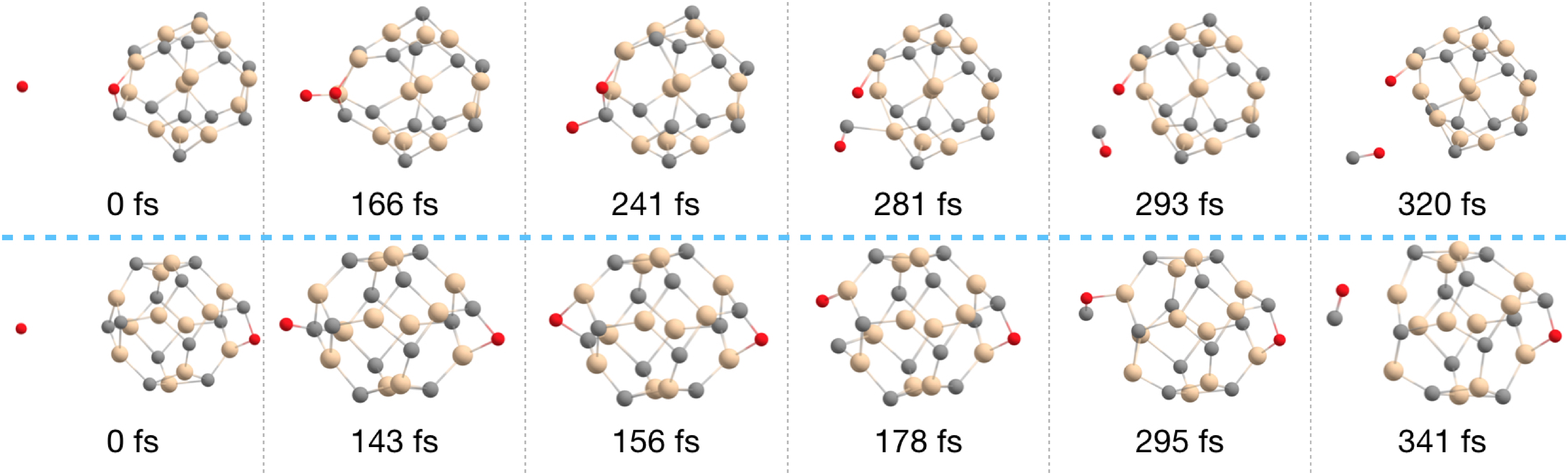}
\caption{%\footnotesize
            \label{fig:snapshot3}
            Snapshots from molecular dynamics simulations
            for collisions between an oxygen atom (red ball)
            and a Si$_{12}$C$_{12}$O molecule
            from two opposite directions.
            }
\end{center}
\end{figure*}

Nevertheless, in the diffuse ISM oxygen atoms
are far more abundant than SiC grains.\footnote{%
  If we take a gas-phase abundance of
  $\ogas=320\ppm$ for atomic oxygen
  in the diffuse ISM (Meyer et al.\ 1998),
  and assume that all SiC grains have the same
  size of $a=0.1\mum$ and consume an upper limit
  of 5$\ppm$ of Si/H (Whittet et al.\ 1990),
  the number density of oxygen atoms
  exceeds that of SiC grains by a factor of
  $\simali$1.3$\times10^{10}$.
%calc 4./3.*3.1416*(0.1e-4)**3*3.22/(40.*1.66e-24)
%=2.03E8 Si atoms
% (320ppm/5ppm)*2.03E8
   }
It is therefore highly probable that a SiC grain
would be hit by multiple oxygen atoms.
Figure~\ref{fig:snapshot3} shows the collisions
of Si$_{12}$C$_{12}$O with another oxygen atom.
The Si$_{12}$C$_{12}$O molecule is adopted from
the simulation illustrated in Figure~\ref{fig:snapshot2} 
but is set to its vibrationally ground state.  
Two opposite incident directions are simulated
for the incoming oxygen atom.
In both cases, we see CO loss within 1\,ps.
This is because, the presence of oxygen atoms
(aka O-substituted molecules) decreases
the transition energy barriers in the region
where the oxygen atom is located,
which makes CO-losses highly favorable
in comparison to other channels.
%e.g., C--H and C=C losses (Chen et al.\ 2018, 2020). 

%Due to the high abundance of hydrogens, besides oxidation,
%hydrogenations are also important for the dissociation of SiC
%molecules. As shown for PAHs, the many additional H atoms (as well as
%for O atoms) in SiC will weaken the backbone structures and increase
%the fragmentation rate
%(Gatchell et al. 2015, Phs. Rev. A, 92, 050702).
%In addition, the reactions are exothermic,
%meaning that the SiC molecules can be excited
%to high vibrational states, in which the isomerization
%play a key role for ``destructing'' SiC molecules --
%they might be converted to more stable products,
%e.g., PAHs, nanotubes or fullerenes
%(Chen et al. 2018, 2019).  

%%% Figure 3 %%%
%%% Figure 3 %%%

%%% Figure 4 %%%
%%% Figure 4 %%%

%%% Figure 5 %%%
%%% Figure 5 %%%

\section{Discussion}\label{sec:discussion}
We have shown in \S\ref{sec:results} that the reaction
between an oxygen atom and a SiC molecule is 
exothermic and would result in CO-loss
almost instantly for small SiC molecules,
or after the absorption of one or several oxygen atoms
for large SiC molecules.
Therefore, the destruction rate of SiC in the ISM
is essentially determined by the collision rate
of oxygen atoms with SiC grains.  
Let $\nO$ be the number density of oxygen atoms
in the diffuse ISM, $\vO$ be the mean thermal velocity
of oxygen atoms, and $\taucoll$ be the collision time
scale of an oxygen atom with a SiC grain of radius $a$.
The collision rate (in $\s^{-1}$) is
$\taucoll^{-1} = \nO\,\vO\,\pi a^2$,
where $\nO\approx\ogas\nH$,
$\vO=\sqrt{3\kB\Tgas/\mO}$,
$\nH$ is the hydrogen number density,
$\kB$ is the Boltzman constant,
$\Tgas$ is the gas temperature,
and $\mO$ is the mass of an oxygen atom.
If we adopt an interstellar gas-phase oxygen abundance
of $\ogas=320\ppm$ (Meyer et al.\ 1998),
$\nH=1\cm^{-3}$ and $\Tgas=100\K$
for the diffuse ISM, we obtain
$\taucoll\approx 2.52\times10^8\times
\left(0.1\mum/a\right)^2\s
\approx7.99\times\left(0.1\mum/a\right)^2\yr$.
%
% vO=3.95E4 cm/s;
%1yr = 3.15E7 s;
%calc 3.2e-4*3.95e4*3.1416*1e-10 = 3.97e-9 s^-1;
%1/3.97e-9 = 2.52e8 s = 7.99 yr;
%calc 4./3.*3.1416*(0.1e-4)**3*3.22 = 1.35e-14 g
%calc 4.9e9*1.989e33*5e-6*40/1.35e-14
%

Let $\sisic$ be the abundance of silicon
(relative to hydrogen) tied up in SiC dust
in the ISM, $\music=40$ be the molecular weight
of SiC, $\rhosic\approx3.2\g\cm^{-3}$
be the mass density of SiC dust,\footnote{%
      We note that such a mass density
      is applicable for bulk, macroscopic
      SiC dust. For small SiC clusters,
      this would underestimate
      the actual ``size'' since small SiC clusters
      are hollow and not compactly packed
      (e.g., such a mass
      density would imply a radius of
      $a\approx3.90\Angstrom$ for
      a spherical Si$_{12}$C$_{12}$ cluster).
    }
$\msic=\left(4/3\right)\pi a^3\rhosic$
be the mass of a spherical SiC grain of radius $a$,
and $\MH$ be the total interstellar
hydrogen gas mass in the Milky Way.
The total number of SiC grains would be
$\Nsic\approx\MH\sisic\music/\msic
\simlt 1.44\times10^{53}\left(0.1\mum/a\right)^3$,
if we adopt $\MH=4.9\times10^9\Msun$
(Draine 2011), $\sisic\simlt5\ppm$ (Whittet et al.\ 1990),
and assume that all SiC grains have the same size $a$.
If each collision of an oxygen atom with a SiC grain
results in an ejection of CO, we would expect a total
SiC dust mass destruction rate of
$\dmdtsicdes\approx12\mH\Nsic/\taucoll
\simlt 1.80\times10^{-4}\left(0.1\mum/a\right)\Msun\yr^{-1}$.

The injection rate of carbon dust from
all carbon-rich objects including
C, R, N, and S stars
into the ISM was estimated to be
$\dmdtcpro\approx0.003-0.01\Msun\yr^{-1}$ (Gehrz 1989).
Lets assume the SiC dust injection rate
from evolved stars to be 10\% of that of carbon dust
(see Groenewegen et al.\ 1998, 2009;
Nanni et al.\ 2013, 2019). We would then expect that
each year carbon stars eject about
$\left(3-10\right)\times10^{-4}\Msun$
of SiC dust into the ISM,
i.e., $\dmdtsicpro\approx10\%\times\dmdtcpro
\approx\left(3-10\right)\times10^{-4}\Msun\yr^{-1}$.
Compared with the SiC destruction rate of
$\dmdtsicdes \simlt1.80\times10^{-4}
\left(0.1\mum/a\right)\Msun\yr^{-1}$,
it is clear that one requires the interstellar SiC grains
to be smaller than $\simali$0.02$\mum$
(for $\dmdtsicpro = 3\times10^{-4}\Msun\yr^{-1}$)
or $\simali$0.06$\mum$
(for $\dmdtsicpro = 1\times10^{-3}\Msun\yr^{-1}$)
in order for the destruction rate to exceed
the current injection rate from carbon stars,
i.e., $\dmdtsicdes \simgt \dmdtsicpro$. 
However, presolar SiC grains of stellar origin
identified in primitive meteorites
are mostly submicron-sized.\footnote{%
  While Amari et al.\ (1994) found that
  $\simali$20\% of all the presolar SiC grains
  seen in the Murchison meteorite is in the fractions
  greater than 1$\mum$ and only about 4\%
  in the fraction less than 0.3$\mum$,
  NanoSIMS measurements of the Murchison meteorite
  with a resolution of $\simali$50\,nm
  revealed that submicron-sized grains
  are much more abundant than their larger,
  micron-sized counterparts (see Hoppe et al.\ 2010).
  }
  %
%Theoretical calculations of SiC condensation
%in evolved stars also point to a size distribution
%peaking at $\simali$0.2--0.3$\mum$,
%consistent with that of presolar SiC grains
%found in meteorites (see Cristallo et al.\ 2020).
%

As we have already generously assumed that
one collision leads to the ejection of one CO molecule
despite that for large SiC molecules multiple collisions
are needed, we conclude that the destruction rate
of SiC dust by oxidation is unlikely larger than the stellar
injection rate and therefore the nondetection of
the 11.3$\mum$ absorption feature of SiC dust
in the diffuse ISM cannot be explained
by the destruction of SiC dust through oxidation.
It is true that the kinetic reactions of oxygen
atoms with bulk, macroscopic SiC dust certainly
differ from that of oxygen atoms with molecular
SiC clusters. However, the destruction rates of SiC
dust derived above do not rely on the exact oxidation
rates of SiC clusters. Indeed, the only assumption
made in deriving $\dmdtsicdes$ is that we have
assumed that each collision of an oxygen atom
with a SiC dust grain would result in an ejection
of CO. As mentioned above, this would result in
an overestimation of $\dmdtsicdes$
since for large SiC clusters a single collision
is unlikely capable of releasing CO.

While the 11.3$\mum$ absorption feature could
be suppressed in large SiC grains of $a>1\mum$,
presolar SiC grains
% and theoretical calculations of SiC dust
% condensation in AGB stars,
as mentioned earlier, have indicated the predominant
presence of submicron-sized SiC grains.
Perhaps submicron-sized SiC grains
only constitute a small fraction of
the SiC dust condensed in carbon stars
and then ejected into the ISM.
Cristallo et al.\ (2020) examined the evolution of the AGB
stars that polluted the solar system at 4.57\,Gyr ago and
found that the submicron-sized presolar SiC grains
predominantly originated from AGB stars with solar metallicity
and an initial mass of $\simali$2$\Msun$.
Theoretical calculations have shown that 
the typical grain size of SiC condensed in AGB stars
can be dependent on the stellar metallicity
and can be smaller than presolar SiC grains
for stars with metallicities lower than solar
(e.g., see Figures~11 and 12 in Nanni et al.\ 2013,
Sect. 5.3 in Ventura et al.\ 2014, Section 5.3).
If the majority of the interstellar SiC grains
originated from AGB stars are smaller than
0.02$\mum$ or even nano-sized,
they will be destroyed more rapidly
than the currently believed stellar injection rate
since the SiC destruction rate
increases as the SiC dust size decreases,
$\dmdtsicdes\propto 1/a$.
In the diffuse ISM, nano-sized SiC grains  
will be stochastically heated by single,
  individual stellar photons (Draine \& Li 2001)
  and will emit at the Si--C stretch,
  presumbly at 11.3$\mum$, which,
  because of the quantum-confinement effect,
  is expected to differ from
  the 11.3$\mum$ absorption feature
  associated with the (crystalline) bulk phase of SiC.
  Like crystalline silicate nanoparticles,
  the 11.3$\mum$ emission feature of
  SiC nanoparticles will probably be hidden
  by that of polycyclic aromatic hydrocarbon
  molecules (see Li \& Draine 2001).

Admittedly, the average mass loss rates and
total numbers for carbon stars in the Galaxy
are not accurately known.
The dust production rates of carbon stars
depend on the star formation history and
metallicity of the galaxy under consideration.
The SiC mass fraction depends on the metallicity as well.
Nanni et al.\ (2019) found that the SiC mass fraction
could be up to 43\% of the total dust mass produced
by the carbon stars in the Large Magellanic Cloud
and 11\% for the Small Magellanic Cloud.
However, these are upper limits reached
by the dustiest carbon stars, and the spreads
in the SiC mass fractions are large
(see Figure~9 of Nanni et al.\ 2019).
As a consequence, the injection rate of SiC dust
into the ISM may be much smaller.
In Groenewegen et al.\ (1998) where the spectral
energy distribution fitting had been performed
for carbon stars in the Galaxy, the SiC mass fractions
are smaller than 10\% for most (37/44) of their stars.
Therefore, the nondetection of the 11.3$\mum$
absorption feature of SiC dust could merely be
due to a lower injection rate of carbon dust
and/or a lower SiC mass fraction
than that adopted here.

Alternatively, as argued by Draine (1990),
interstellar dust is {\it not} stardust,
i.e., the bulk of the solid material in interstellar grains
actually condensed in the ISM rather than in stellar outflows.
While carbon stars do eject an appreciable amount
of SiC dust into the ISM, supernova shock waves destroy
SiC dust (as well as silicate and carbon dust) at a rate faster
than its production (McKee 1989), while grain re-growth
in the oxygen-rich ISM unlikely leads to SiC.

Finally, we note that the B3LYP hybrid functional
employed here had been shown to outperform
for SiC cluster systems (see Byrd et al.\ 2016).
While the triangle isomer of Si$_3$C$_3$
and the spherical isomer Si$_{12}$C$_{12}$
adopted here correspond to their respective global minimum
based on DFT calculations made at the M11/cc-pvTZ level
of theory (Gobrecht et al.\ 2017), a different global minimum
which is lower than the spherical isomer by $>$\,0.4\,eV
was found for Si$_{12}$C$_{12}$ from calculations
at the B3LYP/6-311++G(2d,p) level (Byrd et al.\ 2016).  
%
% Moreover, Byrd et al 2016 found a different global minimum
% for Si12C12 (denoted as closo) than the spherical isomer
% used in this study. Compared to the spherical isomer,
% the closo Si12C12 structure is significantly lower in energy
% (by more than 0.40 eV) using the level of theory
% applied by the authors (B3LYP/6-311++G(2d,p)
% including Grimmes dispersion including Becke-Johnson damping).
% Also for Si3C3 I found several lower-energy isomers
%applying the methods used by the authors.
%Therefore, I kindly ask the authors to either use
%the lowest-energy isomers using the B3LYP/6-311++G(2d,p)
%calculations (that are different from the structures presented
%in this study) and to indicate that the B3LYP functional is not
%very accurate for Si-C systems, or, to apply a different
%functional/basis set, in which the triangular Si3C3 structure
%and spherical Si12C12 structure corresponds to their respective
%global minimum.
%
In view of this inaccuracy, we have also calculated the binding
energies for the O\,+\,Si$_3$C$_3$ system
and the O\,+\,Si$_{12}$C$_{12}$ system
using the M11/cc-pvTZ level of theory
applied by Byrd et al.\ (2016).
It is found that the energy difference
between  M11/cc-pvTZ and B3LYP/6-311++G(2d,p)
is at most 0.24\,eV for O\,+\,Si$_{3}$C$_{3}$ and
only $\simali$0.01\,eV for O\,+\,Si$_{12}$C$_{12}$.
Such energy differences will not affect the major
results of this work.

\section{Conclusion}\label{sec:conclusion}
We have utilized molecular dynamics simulations
and performed DFT calculations to investigate
the oxidation of SiC dust in the ISM.
It is found that, although the reaction of an oxygen
atom with a SiC molecule is exothermic and could
cause CO-loss, the destruction rate of SiC dust
through oxidation is considerably smaller than 
the currently believed stellar injection rate and
therefore the nondetection of the 11.3$\mum$
absorption feature of SiC dust in the diffuse ISM
cannot be explained by the destruction of SiC dust
through oxidation,
unless the currently believed SiC dust injection
rate from carbon stars is overestimated and/or
interstellar SiC dust is considerably smaller than
submicron in size.

\section*{Acknowledgements}
The calculations were performed on resources
provided by the Swedish National Infrastructure
for Computing (SNIC).
We thank D.~Gobrecht, M.A.T.~Groenewegen,
P.F.~Miceli, A.~Nanni, D.J.~Singh, X.J.~Yang
and the anonymous referees
for very helpful suggestions.
CYX is supported in part by 
the Talents Recruiting Program of Beijing Normal University
and the NSFC Grant No.\,91952111.
A.L. is supported in part by 
NSF AST-1816411 and NASA 80NSSC19K0572. 

\section*{Data Availability}
The data underlying this article will be shared 
on reasonable request to the corresponding authors.

\bsp
\label{lastpage}


\begin{thebibliography}{}
  \expandafter\ifx\csname natexlab\endcsname\relax\def\natexlab#1{#1}\fi

\bibitem[]{}Amari, S., Lewis, R. S.,
                  \& Anders, E.\ 1994,
                  Geochim. Cosmochim. Acta, 58, 459

% \bibitem[]{}Amari, S., Zinner, E., \& Gallino, R.\
%                  2014, AIP Conf. Ser. 1594,
%                  Origin of Matter and Evolution of Galaxies,
%                  ed. S. Jeong, et al.\ (Melville, NY: AIP), 307

\bibitem[{Becke(1992)}]{becke1992density}
  Becke, A.~D.\ 1992, J. Chem. Phys., 96, 2155
  
\bibitem[{Bernatowicz {et~al.}(1987)Bernatowicz, Fraundorf, Ming, Anders,
    Wopenka, Zinner, \& Fraundorf}]{bernatowicz1987evidence}
  Bernatowicz, T., Fraundorf, G., Ming, T., {et~al.} 1987, Nature, 330, 728

\bibitem[]{}Byrd, J.N., Lutz, J.J., Jin, Y.F., et al.\ 2016,
                  J. Chem. Phys., 145, 024312

%\bibitem[]{}Chen, T.


\bibitem[]{}Cristallo, S., Nanni, A., Cescutti, G., et al.\ 2020, A\&A, 644, A8 

%  \bibitem[{Daulton {et~al.}(2003)Daulton, Bernatowicz, Lewis, Messenger,
%    Stadermann, \& Amari}]{daulton2003polytype}
%  Daulton, T., Bernatowicz, T., Lewis, R., {et~al.} 2003, Geochimica et
%    Cosmochimica Acta, 67, 4743
  
\bibitem[{Dorfi \& H{\"o}fner(1991)}]{dorfi1991dust}
  Dorfi, E., \& H{\"o}fner, S.\ 1991, A\&A, 248, 105

\bibitem[]{}Draine, B.T.\ 1990, in ASP Conf. Ser. 12, The Evolution of 
       the Interstellar Medium, ed. L. Blitz (San Francisco: ASP), 193

\bibitem[]{}Draine, B.T.\ 2011, Physics of the Interstellar
                  and Intergalactic Medium
                 (Princeton, NJ: Princeton Univ. Press)

\bibitem[]{}Draine, B.T., \& Li, A.\ 2001, ApJ, 551, 807
                 
\bibitem[]{}Ervin, G.\ 1958, J. Am. Ceram. Soc., 41, 347 
                 
\bibitem[]{}Friedemann, C.\ 1969, Astron. Nachr., 291, 177

% \bibitem[{Frisch {et~al.}(2016)Frisch, Trucks, Schlegel, Scuseria, \&
%  et~al.}]{frisch2016gaussian}
    

\bibitem{}Frisch, M. J., Trucks, G. W., Schlegel, H. B.,
            et al.\ 2016, Gaussian 16, Revision C. 01,
            Gaussian, Inc., Wallingford CT

\bibitem[]{}Gehrz, R.D.\ 1989, in IAU Symp. 135,
                  Interstellar Dust,  ed. L. J. Allamandola
                  \& A. G. G. M. Tielens (Dordrecht: Kluwer), 445
            
\bibitem[]{}Gilman, R.C.\ 1969, ApJ, 155, L185
  
\bibitem[]{}Gobrecht, D., Cristallo, S., Piersanti, L.,
                 \& Bromley, S.~T.\ 2017, ApJ, 840, 117 

\bibitem[]{}Grimme, S., Ehrlich, S., \& Goerigk, L.\
                  2011, J. Comput. Chem., 32, 1456
                 
\bibitem[]{}Groenewegen, M.~A.~T., Whitelock, P.~A.,
                  Smith, C.~H., et al.\ 1998, MNRAS, 293, 18
  
\bibitem[]{}Groenewegen, M.~A.~T., Sloan, G.~C., Soszy{\'n}ski, I.,
                  et al.\ 2009, A\&A, 506, 1277
  
\bibitem[]{}Helgaker, T., Uggerud, E., \& Jensen, H.J.A.\
                  1990, Chem. Phys. Lett., 173, 145

%\bibitem[{Hoppe {et~al.}(1996)Hoppe, Strebel, Eberhardt, Amari, \&
%    Lewis}]{hoppe1996small}
%  Hoppe, P., Strebel, R., Eberhardt, P., Amari, S., \& Lewis, R.~S. 1996,
%    Geochimica et cosmochimica acta, 60, 883

\bibitem[]{}Hoppe, P., Leitner, J., Gr\"oner, E.\
                  et al.\ 2010, ApJ, 719, 1370
    
%\bibitem[{Jones {et~al.}(1978)Jones, Merrill, Puetter, \&
%   Willner}]{jones1978infrared}
%  Jones, B., Merrill, K., Puetter, R., \& Willner, S. 1978, The Astronomical
%    Journal, 83, 1437
  
%  \bibitem[{Jura \& Kleinmann(1989)}]{jura1989dust}
%  Jura, M., \& Kleinmann, S. 1989, The Astrophysical Journal, 341, 359

\bibitem[]{}Kraemer, K.~E., Sloan, G.~C., Keller, L.~D., et al.\ 2019, ApJ, 887, 82

\bibitem[]{}Larsson, B., Liseau, R., Pagani, L.,
                  et al.\ 2007, A\&A, 466, 999
  
\bibitem[{Lee {et~al.}(1988)Lee, Yang, \& Parr}]{lee1988development}
  Lee, C., Yang, W., \& Parr, R.~G.\ 1988, Phys. Rev. B, 37, 785

\bibitem[]{}Li, A.\, \& Draine, B.T.\ 2001, ApJ, 550, L213
  
\bibitem[]{}McKee, C.F.\ 1989, in IAU Symp. 135, Interstellar Dust, 
ed. L. J. Allamandola \& A. G. G. M. Tielens (Dordrecht: Kluwer), 431


\bibitem[Meyer et al.(1998)]{Meyer1998}
Meyer, D.M., Jura, M., \& Cardelli, J.A.\ 1998, ApJ, 493, 222

\bibitem[]{}M\"uhlh\"auser, M., Froudakis, G., Zdetsis, A.,
                  \& Peyerimhoff, S.D.\ 1993, Chem. Phys. Lett., 204, 617

\bibitem[]{}Mutschke, H., Andersen, A.~C., Cl{\'e}ment, D.,
                  Henning, Th., \& Peiter, G.\ 1999, A\&A, 345, 187

\bibitem[]{}Nanni, A., Bressan, A., Marigo, P., et al.\ 2013, MNRAS, 434, 2390

\bibitem[]{}Nanni, A., Bressan, A., Marigo, P., et al.\ 2014, MNRAS, 438, 2328
  
\bibitem[]{}Nanni, A., Groenewegen, M.~A.~T., Aringer, B., et al.\ 2019, MNRAS, 487, 502

\bibitem[]{}Nanni, A., Cristallo, S., van Loon, J.~T., et al.\ 2021, Universe, 7, 233

\bibitem[]{}Park, D.J., Jung, Y.I., Kim, H.G., Park, J.Y.,
                 \& Koo, Y.H.\ 2014, Corros. Sci., 88, 416
  
\bibitem[]{}Sloan, G.~C., Kraemer, K.~E., McDonald, I., et al.\ 2016, ApJ, 826, 44
  
%  \bibitem[{Speck {et~al.}(2006)Speck, Cami, Markwick-Kemper, Leisenring,
%    Szczerba, Dijkstra, Van~Dyk, \& Meixner}]{speck2006unusual}
%  Speck, A.~K., Cami, J., Markwick-Kemper, C., {et~al.} 2006, The Astrophysical
%    Journal, 650, 892

\bibitem[]{}Speck, A.~K., Barlow, M.~J., \& Skinner, C.~J.\ 1997, MNRAS, 288, 431

  
%  \bibitem[{Stroud \& Bernatowicz(2005)}]{stroud2005surface}
%  Stroud, R., \& Bernatowicz, T.~J. 2005, Lunar and Planetary Science, 36
  
\bibitem[{Treffers \& Cohen(1974)}]{treffers1974high}
  Treffers, R., \& Cohen, M.\ 1974, ApJ, 188, 545
  
\bibitem[{Uggerud \& Helgaker(1992)}]{uggerud1992dynamics}
  Uggerud, E., \& Helgaker, T.\ 1992, J. Am. Chem. Soc.,
    114, 4265
  
\bibitem[{Van~Loon {et~al.}(2005)Van~Loon, Cioni, Zijlstra, \&
    Loup}]{van2005empirical}
  van~Loon, J.~T., Cioni, M.-R., Zijlstra, A.~A., \& Loup, C.\ 2005, A\&A, 438, 273


\bibitem[]{}Ventura, P., di Criscienzo, M., Schneider, R.,
                  et al.\ 2012, MNRAS, 424, 2345

\bibitem[]{}Ventura, P., Dell'Agli, F., Schneider, R.,
                  et al.\ 2014, MNRAS, 439, 977
  
\bibitem[]{}Ventura, P., Karakas, A., Dell'Agli, F.,
                 Garc{\'i}a-Hern{\'a}ndez, D.A.,
                 Guzman-Ramirez, L.\  2018, MNRAS, 475, 2282

\bibitem[]{}Watkins, M.~B., Shevlin, S.~A., Sokol, A.~A.,
                 et al.\ 2009, Phys. Chem. Chem. Phys., 11, 3186
                 
\bibitem[{Whittet {et~al.}(1990)Whittet, Duley, \&
    Martin}]{whittet1990abundance}
  Whittet, D.C.B., Duley, W.W., \& Martin, P.G.\ 1990, MNRAS, 244, 427

\bibitem[]{}Wildt, R.\ 1933, Zeitschr. f\"ur Astrophys., 6, 345
  
\bibitem[]{}Zhukovska, S., \& Henning, Th.\ 2013, A\&A, 555, A99
    
\end{thebibliography}
\end{document}